\documentclass{article}

\usepackage[autostyle=true]{csquotes}
\usepackage{tikz}
\usepackage[margin=1.1in]{geometry}
\usepackage{url}
\usepackage{soul,xcolor,amsmath,amssymb}
\usepackage{graphicx}
\usepackage{lineno}
\usepackage{textcomp}
\usepackage{booktabs}
\usetikzlibrary{calc}
\usetikzlibrary{3d}
\usepackage{bm}
\usepackage{caption}
\usepackage{caption}
\usepackage{subcaption}
\usepackage{float}
\usepackage{authblk}

\begin{document}

\setstcolor{red}


\newcommand{\Einc}{E^{\rm inc}}
\newcommand{\Esca}{E^{\rm sca}} 
\newcommand{\Etot}{E^{\rm tot}}
\newcommand{\EL}{E_L}
\newcommand{\x}{\bm{x}}
\newcommand{\y}{\bm{y}}
\newcommand{\Rtwo}{\bm{R}^2}

\title{Programmable photonic nanojets via phase-only time-reversal:\\ a numerical study}

\author[1]{Tobias Abilock Mikkelsen}
\author[1]{Cristian Placinta}
\author[2]{Jesper Glückstad}
\author[1]{Mirza Karamehmedovi\'c\thanks{\texttt{mika@dtu.dk}}}
\affil[1]{Department of Applied Mathematics and Computer Science, Technical University of Denmark, DK-2800 Kgs. Lyngby, Denmark}
\affil[2]{SDU Centre for Photonics Engineering, University of Southern Denmark, DK-5230 Odense, Denmark}

\date{}

\maketitle

\abstract
We present a phase-only time-reversal framework for steering photonic nanojets without mechanical motion or amplitude modulation. Time-reversed radiation by a synthetic source placed at the target PNJ location helps define a phase-only modulation on a control line, compatible with a spatial light modulator, that produces the desired PNJ. Full-wave finite-difference frequency-domain (FDFD) simulations demonstrate robust lateral and axial steering with subwavelength confinement and low sidelobes. A parametric study of microelement geometries shows that nanojet formation is largely insensitive to moderate boundary variations, with simple shapes providing competitive performance. Robustness to fabrication and alignment errors is confirmed via uncertainty analysis.

\section{Introduction}

Photonic nanojets (PNJs)~\cite{Darafsheh-2021, Lecler-2019, Itagi-2005, 2023-phase-only_PNJ, 2022-PNJ1} are highly concentrated, high-intensity optical structures that form on the shadow side of illuminated mesoscale dielectrics. They have enabled super-resolution imaging~\cite{Huszka-2019}, nanopatterning~\cite{Zelgowski-2016,Zhang-2013}, optical manipulation~\cite{Rodrigo2}, and label-free metrology~\cite{Lai-2016,2022-SPIE} while avoiding near-field probes and plasmonic losses. A long-standing obstacle to broader deployment is agile, calibration-friendly control of PNJ position and morphology without mechanical motion. Existing approaches (see, e.g.,~\cite{2023-phase-only_PNJ} for a comprehensive literature list) are based on microelement shaping (geometry, materials, arrays, substrates) or illumination beam shaping (beam focus, waist radius, and axis control). While these strategies can steer field confinement, they are typically slow, offer only modest steering ranges, or require complex hardware and multi-parameter calibrations. In our prior work we laid the foundations for software-defined PNJ control: first, we developed an inverse-scattering formulation (via a Lippmann--Schwinger representation) that computes the complex incident field required to realize a target near-field PNJ at prescribed locations, establishing a predictive route to model-based steering \cite{2022-PNJ1}; and second, we demonstrated hardware-lean phase-only illumination for axial and angular PNJ control using a single spatial light modulator (SLM), together with a practical calibration that improved confinement while preserving optical throughput \cite{2023-phase-only_PNJ}. See also our related work in~\cite{Mogensen,GluckstadMadsen-2023}.

We here present a phase-only illumination modulation framework that achieves wide-range PNJ steering for general dielectric microelements using a time-reversal numerical scheme. Given a desired PNJ location, a synthetic source produces a scattered field, which in turn is time-reversed and whose phase is sampled over a control surface. The phase, directly displayable on a single SLM, then defines the illumination used to achieve the desired PNJ. The result is a lossless setup that achieves PNJ confinement and low sidelobes across a broad lateral/axial range, and requires no amplitude modulation or moving parts. The setup uses one SLM and a fixed mesoscale microelement, potentially suitable for microscopy, nanofabrication, and trapping systems.

In this paper we restrict attention to a two-dimensional model, which retains the governing physics while enabling computationally expensive quantification metrics requiring large number of iterations. The framework generalizes in principle directly to 3D, using a suitably equipped solver.

Section~\ref{sec:trm} describes the time-reversal method that produces the input phase modulation for a desired PNJ. In Section~\ref{sec:numsim} we detail our numerical scattering solver, as well as the implemented automatic PNJ detection and size measurement. We present numerical results, including uncertainty quantification results, in Section~\ref{sec:numres}, and offer our conclusions in Section~\ref{sec:conclusion}.

\begin{figure}[H]
    \centering
    \includegraphics[width=.5\textwidth]
    {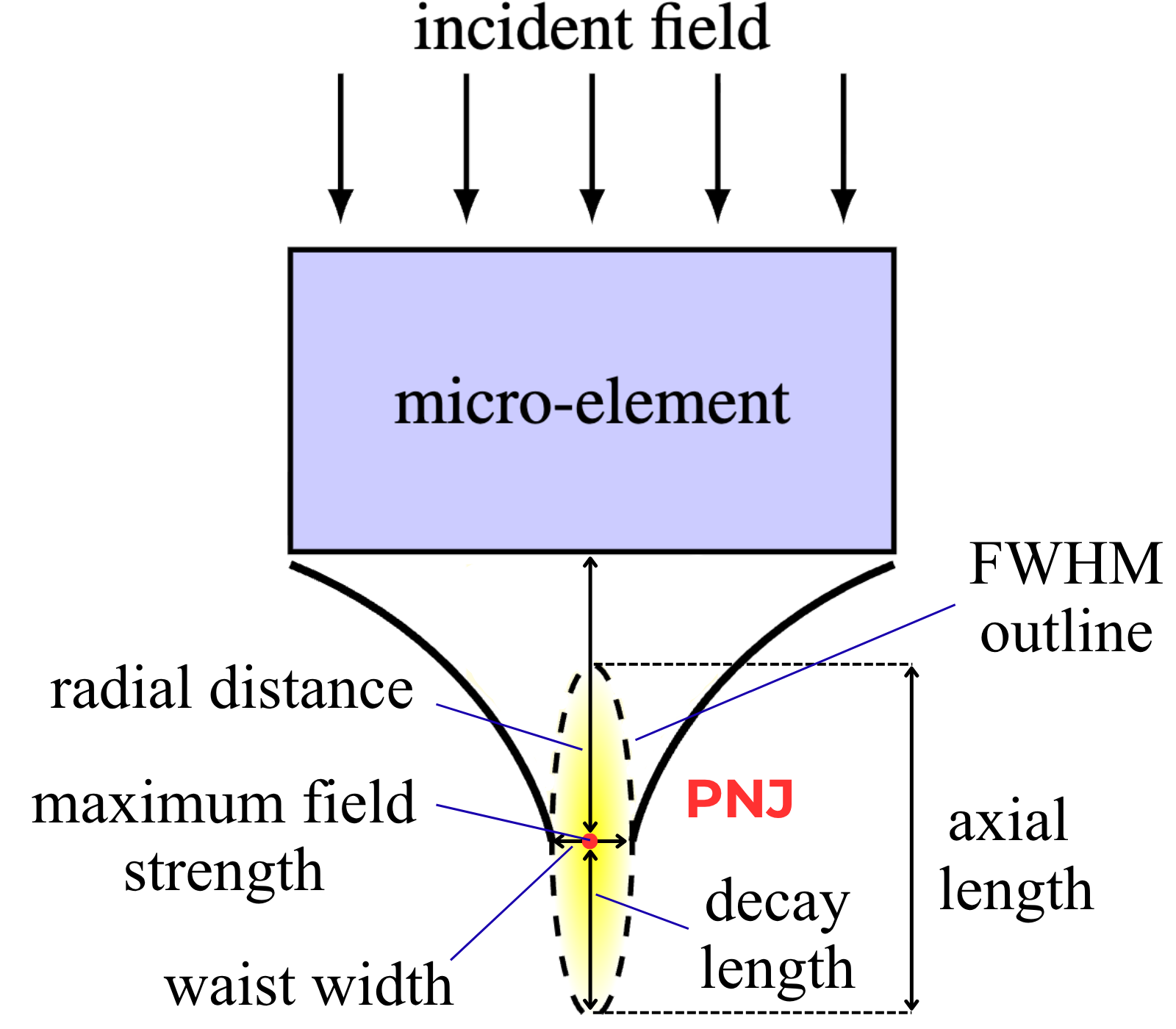}

    \caption{Illuminating a penetrable microelement can result in a highly localized near field (a photonic nanojet, PNJ) on the shadow side. The waist width $w_{\rm PNJ}$ and axial length $l_{\rm PNJ}$ of a PNJ are often measured using the full width at half maximum (FWHM) contour for the total electric field amplitude $|\bm{E}^{\rm tot}|$.}
    \label{fig:PNJsetup}
\end{figure}

\section{The time-reversal method}\label{sec:trm}

\begin{figure}
    \centering\hspace*{1.5cm}
    \includegraphics[width=0.5\linewidth]{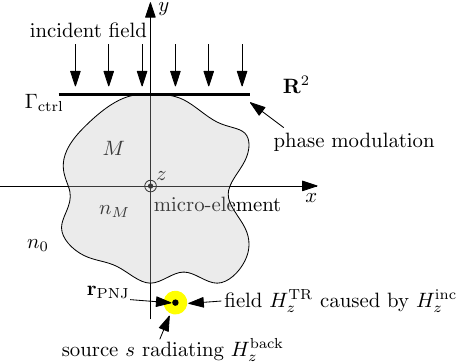}
    \caption{The setup for the time-reversal method.}
    \label{fig:setup}
\end{figure}

The purpose of the time-reversal approach is to compute a phase-only modulation of the incident field that produces a prescribed PNJ in the total near field; see Fig.~\ref{fig:setup}.
We consider two-dimensional transverse electric (2D TE$^z$) time-harmonic scattering, at free-space wavelength $\lambda_0$, by a penetrable, lossless, non-magnetic microelement $M$ with real refractive index $n_M$. We use the convention $e^{-i\omega t}$ in the following. In the first step of the time-reversal method, the microelement is illuminated by a source $s(\bm r)$ concentrated at the target PNJ location $\bm r_{\rm PNJ}$. The total magnetic field $\bm H^{\rm back}(\bm r)=\hat{\bm z}H^{\rm back}_z(\bm r)$ satisfies the equation
\begin{equation}\label{eqn:H1}
\nabla\cdot\left(\frac{1}{\varepsilon_r(\bm r)}\nabla H_z^{\rm back}(\bm r)\right)+k_0^2H_z^{\rm back}(\bm r) = s(\bm r),
\end{equation}
where the wavenumber $k(\bm r)=\omega\sqrt{\mu_0 \varepsilon(\bm r)}$, with the electric permittivity $\varepsilon=\varepsilon_0$ outside $M$ and $\varepsilon=\varepsilon_M=n_M^2\varepsilon_0$ inside. Since the Helmholtz operator has real-valued coefficients, it is invariant under complex conjugation, and~\eqref{eqn:H1} is equivalent with
\begin{equation}
\nabla\cdot\left(\frac{1}{\varepsilon_r(\bm r)}\nabla(H_z^{\rm back})^{\ast}(\bm r)\right)+k_0^2(H_z^{\rm back})^{\ast}(\bm r) = s^{\ast}(\bm r),
\end{equation}
In our implementation, we use the synthetic source
\begin{equation}
s(\bm r) = \exp\!\left(-\frac{|\bm r - \bm r_{\mathrm{PNJ}}|^2}{2\sigma^2}\right),
\end{equation}
with $\sigma$ equal to the width of 3 grid cells, and $\|s\|_\infty=1$. Solving Eq.~\eqref{eqn:H1} using a finite-difference frequency-domain (FDFD) method with perfectly matched layers yields the field $H_z^{\mathrm{back}}(\bm r)$. For the second step, we introduce the control line
\begin{equation}
\Gamma_{\mathrm{ctrl}} = \{(x,y)\in\bm{R}^2 : |x|\le X,\; y=y_{\mathrm{ctrl}}\},
\end{equation}
of length $2X$, located just above the microelement, and spanning its full lateral extent. We decompose $H_z^{\rm back}$ at $\Gamma_{\rm ctrl}$ in terms of real-valued amplitude and phase functions $A(x)$ and $\phi(x)$, $H_z^{\mathrm{back}}(x,y_{\mathrm{ctrl}}) = A(x)e^{i\phi(x)}$. By conjugation invariance, $A(x)e^{-i\phi(x)}$ represents the time-reversed boundary data. In a full time-reversal setting one would impose this field directly. Here, consistent with a phase-only implementation, we enforce a constant amplitude and retain only the conjugated phase,
\begin{equation}
H_z^{\mathrm{inc}}(x,y_{\rm ctrl}) = e^{-i\phi(x)},\quad |x|\le X.
\end{equation}
We now solve the scattering problem with a current sheet injected at $\Gamma_{\rm ctrl}$:
\begin{equation}
\nabla\cdot\left(\frac{1}{\varepsilon_r(\bm r)}\nabla H_z^{\rm TR}(\bm r)\right) + k_0^2 H_z^{\mathrm{TR}}(\bm r) = \delta(y-y_{\rm ctrl})\chi_{\Gamma_{\rm ctrl}}(x)H_z^{\rm inc},
\end{equation}
where $\delta$ is the Dirac delta, and $\chi_{\rm ctrl}(x)=1$ for $x\in\Gamma_{\rm ctrl}$ and $\chi_{\rm ctrl}(x)=0$ elsewhere. By construction, $H_z^{\mathrm{TR}}$ approximates the time-reversed propagation of $H_z^{\mathrm{back}}$ and concentrates energy at $\bm r_{\mathrm{PNJ}}$. The resulting intensity $|H_z^{\mathrm{TR}}(\bm r)|^2$ exhibits a localized, high-contrast maximum at the prescribed location, forming the desired PNJ. The phase profile $e^{-i\phi(x)}$ is obtained directly from the backward simulation and therefore incorporates all refraction, diffraction, and near-field effects within the microelement.

\section{The numerical simulation method}\label{sec:numsim}

All simulations are performed for lossless, nonmagnetic dielectric microelements, at the free-space wavelength \( \lambda_0 = 532\,\mathrm{nm} \). The ambient medium is vacuum with refractive index \( n_0 = 1 \), while the microelements are assigned the real refractive index \( n_{\mathrm{M}} = 1.49 \). The computational domain is a \(10\,\mu\mathrm{m} \times 18\,\mu\mathrm{m}\) rectangular region discretized on a uniform Cartesian grid with a resolution of 30 points per micrometer (ppum), corresponding to approximately 16 points per wavelength. This resolution was verified to provide sufficient accuracy for resolving the field variations. The domain is surrounded by a perfectly matched layer (PML) of thickness \(2\,\mu\mathrm{m}\), ensuring effective absorption of outgoing waves and emulation of an open domain.

The Helmholtz equation is discretized using second-order finite differences, resulting in a sparse linear system $\mathbf{A}\,\mathbf{h} = \mathbf{s}$. Here, \(\mathbf{A}\) represents the discrete Helmholtz operator incorporating the spatially varying relative permittivity \(\varepsilon_r(\mathbf{r})\) as well as the PML boundary terms. The magnetic field \(H_z(\mathbf{r})\) and the source distribution \(S(\mathbf{r})\) are sampled on the grid to form the vectors \(\mathbf{h}\) and \(\mathbf{s}\), respectively. The system is assembled and solved using a full-wave finite-difference frequency-domain (FDFD) formulation implemented in Python via \texttt{ceviche}~\cite{Ceviche}. The excitation is confined to a one-dimensional line source, modeling an equivalent current sheet that launches the incident field. After solving for \(\mathbf{h}\), the in-plane electric field components \(E_x\) and \(E_y\) are reconstructed from Maxwell’s equations. The resulting field intensity is evaluated as \(I(\mathbf{r}) \propto |H_z(\mathbf{r})|^2\), which is used to characterize the photonic nanojet.

\textbf{PNJ quality metrics.} To enable a consistent and physically meaningful characterization of nanojet quality, it is necessary to determine the principal axis of a given PNJ. We define this axis using the local electromagnetic power flow, as described by the time-averaged Poynting vector. In a lossless, nonmagnetic 2D TE setting, the Poynting vector is given by
\[
\bm{S}(x,y) \;=\;
\begin{pmatrix}
\frac{1}{2}\,\Re\!\left\{E_y H_z^{*}\right\} \\
-\frac{1}{2}\,\Re\!\left\{E_x H_z^{*}\right\}
\end{pmatrix}.
\]
Evaluating \(\bm{S}\) at the target nanojet location \(\bm{r}_{\mathrm{PNJ}}\), we define the local power-flow direction
\[
\widehat{\bm{a}} \;=\; \frac{\bm{S}(\bm{r}_{\mathrm{PNJ}})}{\left\lVert \bm{S}(\bm{r}_{\mathrm{PNJ}}) \right\rVert},
\]
which we take as the principal axis of the nanojet. The corresponding transverse direction is defined as the orthogonal unit vector $\widehat{\bm{p}} \;=\; (a_y,\,-a_x)$. The PNJ axis \(\widehat{\bm a}\) and its transverse direction \(\widehat{\bm p}\) are used to define the axial length \(\ell_{\rm PNJ}\) and waist width \(w_{\rm PNJ}\) via local full width at half maximum (FWHM) conditions. Specifically, these are determined from the intensity profile along the axial and transverse directions as
\[
I\bigl(\bm r_{\rm PNJ} + s\,\widehat{\bm a}\bigr) = \tfrac{1}{2}\,I(\bm r_{\rm PNJ}), 
\qquad
I\bigl(\bm r_{\rm PNJ} + t\,\widehat{\bm p}\bigr) = \tfrac{1}{2}\,I(\bm r_{\rm PNJ}),
\]
with \(s = s_L, s_R\) and \(t = t_L, t_R\) denoting the two intersection points on each line. This definition anchors the measurement to the nanojet core (neck) at target \(\bm r_{\rm PNJ}\), rather than to a global intensity maximum, thereby providing a local and geometry-independent characterization of the PNJ. Numerically, starting from a minimal sampling window centered at \(\bm r_{\rm PNJ}\), the window is expanded adaptively along each direction until two crossings of the half-maximum level are detected. In cases where no such crossings are found within a prescribed maximal extent (e.g., due to asymmetry or elevated background levels) a fallback threshold
\[
I_{1/2} \;=\; \tfrac{1}{2}\,\max I_{\parallel,\perp}
\]
is used along the corresponding axial or transverse cut to ensure robustness. The axial length and waist width are then defined as
\[
\ell_{\rm PNJ} = s_R - s_L,
\qquad
w_{\rm PNJ} = t_R - t_L,
\]
corresponding to the axial and transverse FWHM, respectively. We define the local PNJ area as enclosed by a contour formed by two circular arcs passing through the four FWHM crossing points.

\section{Numerical results}\label{sec:numres}

\begin{figure}
    \centering
    \includegraphics[width=1.0\textwidth]{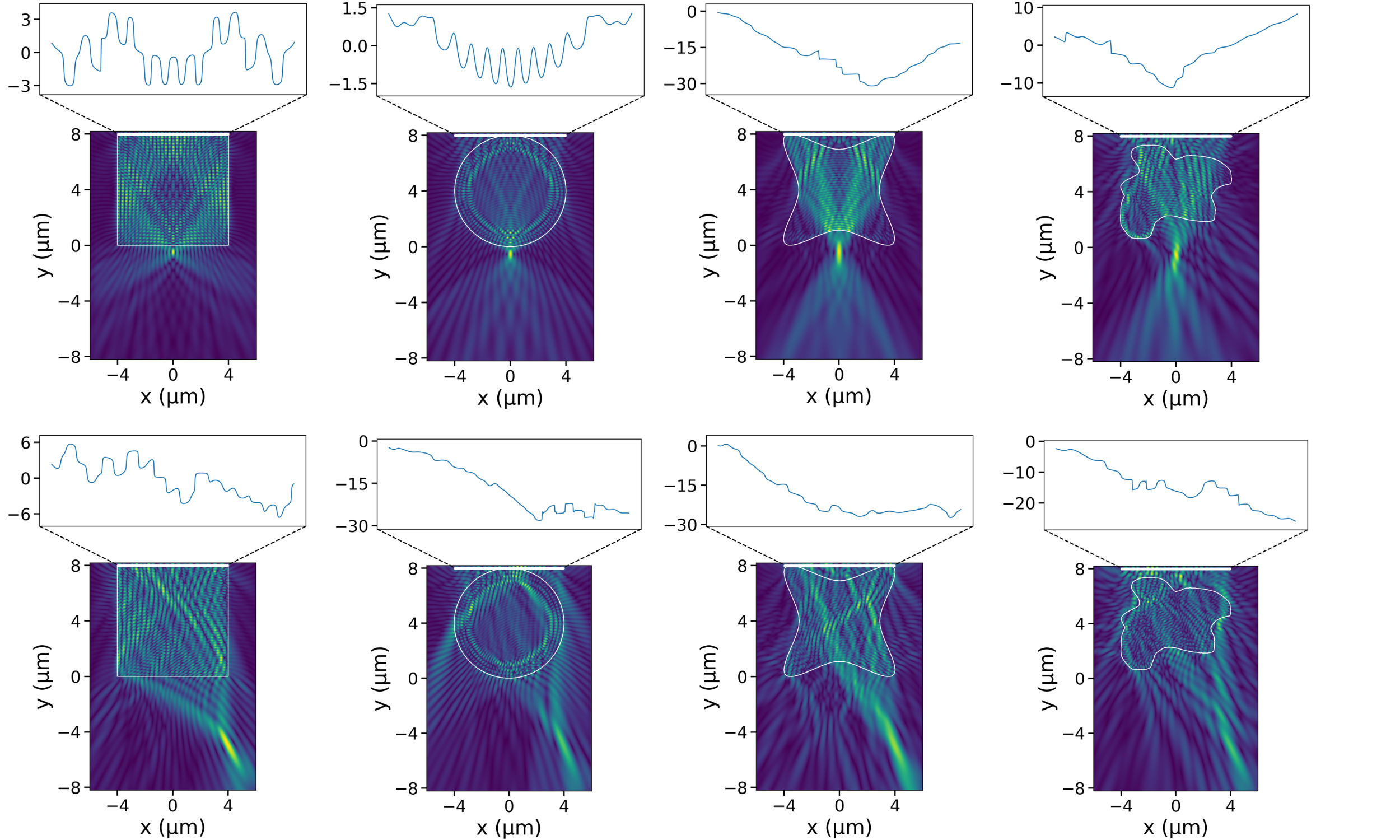} 
    \caption{PNJ steering for four microelement geometries, showing the applied unwrapped input phase profiles. Top: Input phase profiles and the resulting fields for a target PNJ at (0 $\mu\mathrm{m}$, -0.5 $\mu\mathrm{m}$). Bottom: Same microelements with input phase profiles for PNJs steered to (4 $\mu\mathrm{m}$, -5 $\mu\mathrm{m}$). Figures are composed of the normalized field inside the microelement and the independently normalized field outside the microelement.}
    \label{fig:pnj_example}
\end{figure}

\textbf{Microelement geometries.} We first demonstrate PNJ steering numerically for a wide range of microelement geometries. For each geometry, the phase profile is calculated, and confinement and peak intensity are observed. The following results (Fig.~\ref{fig:pnj_example}) confirm that while certain microelements allow higher field confinement, phase-only modulation still enables robust spatial control of the PNJ position independently of the underlying microelement shape. Also, as will become apparent, our broad search here suggests that the method is not highly sensitive to geometry, and that simple high-symmetry shapes are already highly competitive.

To quantify the influence of geometry on PNJ performance, we introduce a constrained version of the superformula as a parameterization that enables systematic exploration of a wide range of microelement shapes:
\begin{equation}\label{eq:superformula}
r(\theta) \;=\;
\left(\;
\left|\,\cos\left(\frac{n\theta}{4}\right)\right|^{m_2}
+
\left|\,\sin\left(\frac{n\theta}{4}\right)\right|^{m_3}
\,\right)^{-1/m_1},\quad\theta\in[0,2\pi).
\end{equation}
We impose $n \in \{2,\dots,20\}$, which controls the number of symmetry lobes
of the boundary, while $m_1$, $m_2$, and $m_3 \in (0,10]$ determine
the sharpness and curvature of these features. In this simulation, we restrict the
parameter space by imposing $m_2 = m_3$, which enforces symmetric curvature and
avoids degenerate or highly anisotropic shapes.

\begin{figure}
    \centering
    \begin{subfigure}{0.47\textwidth}
        \includegraphics[width=\textwidth]{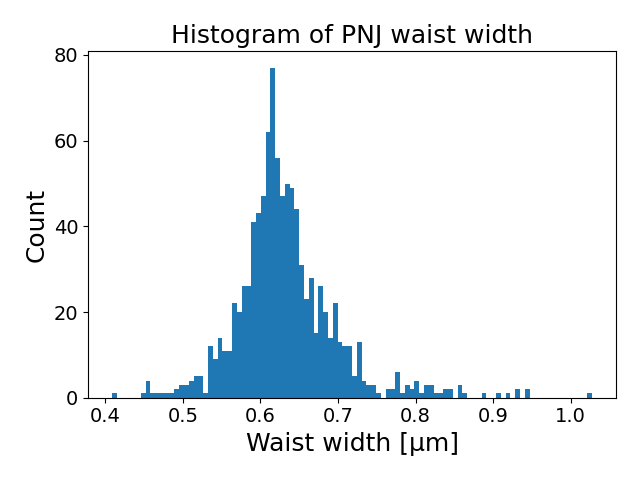}
    \end{subfigure}
    \hfill
    \begin{subfigure}{0.47\textwidth}
        \includegraphics[width=\textwidth]{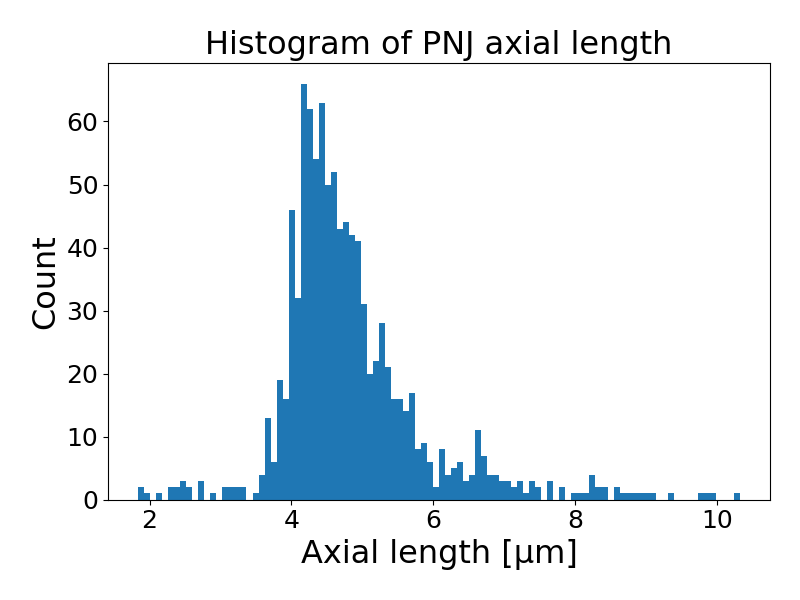}
    \end{subfigure}
    \caption{Waist width and axial length histograms of 1000 randomly sampled superformula microelement shapes \cite{Thesis}.}
    \label{fig:histograms_superformula}
\end{figure}

Using the superformula parametrization, we generate $1000$ uniformly random microelement
geometries and compute the corresponding PNJs for a fixed target
position at $(0,-4)\,\mu\mathrm{m}$. This random sampling enables an unbiased
exploration of the geometric design space and provides a statistical view of
how arbitrary microelement shapes influence nanojet formation. For each geometry, the nanojet waist width and
axial length are extracted. Despite the lack of structure in the selection process, the resulting distribution of nanojet waist widths is approximately Gaussian. In contrast, the axial length histogram exhibits a noticeable right skew, indicating increased sensitivity to elongated axial profiles (Fig.~\ref{fig:histograms_superformula}). This behavior indicates that nanojet formation under time-reversal illumination is relatively insensitive to moderate boundary perturbations.

Across the sampled natural geometries, the resulting nanojet quality metrics cluster within a relatively narrow range, indicating that most microelement geometries yield comparable nanojet performance, with only a limited number of strongly deviating cases. This statistical baseline is useful for optimization, as it provides a reference against which improvements from targeted geometric design can be evaluated.

To compare geometries quantitatively, we introduce a composite performance
metric that favors high intensity together with strong transverse confinement and
short axial extent,
\begin{equation}\label{eq:score}
\text{score} \;=\;
\frac{I_{\text{avg}}}{w_{\rm PNJ}\cdot\ell_{\rm PNJ}} \;,\;\; \text{where $I_{\text{avg}}$ denotes the average intensity over PNJ area.}
\end{equation}

\begin{figure}[H]
    \centering
    \includegraphics[width=.87\textwidth]{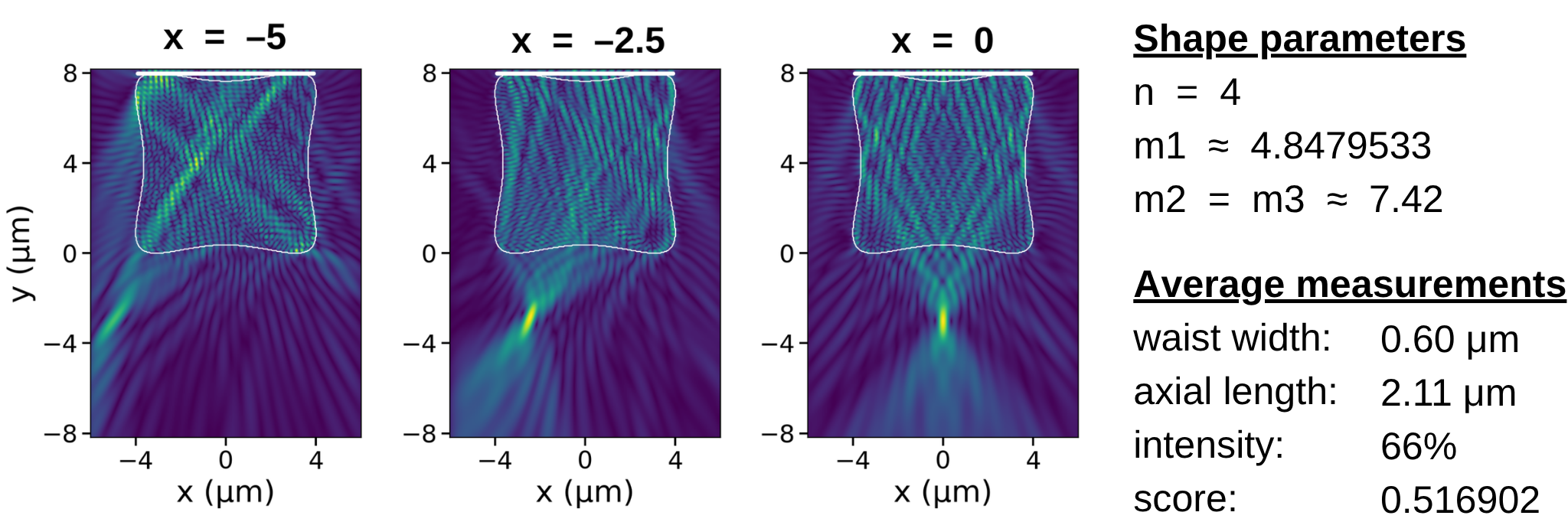}    
    \caption{Time-reversal field intensities produced by the selected best-performing superformula microelement shape for different target nanojet positions. The right panel reports the corresponding shape parameters and averaged nanojet performance metrics over all evaluated targets \cite{Thesis}. }
    \label{fig:best_lens}
\end{figure}

Using this measure, we select the $20$ best-performing geometries from the
simulations. Interestingly, the selected shapes consistently exhibit four
dominant corners, suggesting that this may be beneficial for efficient nanojet formation. We then perform a second parameter sweep restricted to superformula shapes with four dominant corners $(n=4)$. An additional $1000$
geometries are generated within this constraint, and the top $20$
candidates are selected according to the composite score. These candidates
are then evaluated across multiple target nanojet positions,
\[
(-5,-3),\; (-2.5,-3),\; (0,-3),\;
(-5,-6),\; (-2.5,-6),\; (0,-6),
\qquad\quad(\text{in } \mu m)
\]
and the final performance score is obtained by averaging over all targets.
Figure~\ref{fig:best_lens} shows the best-performing geometry identified in this
restricted search.

While the superformula search enables exploration of a flexible family of natural geometries, it is also important to compare these optimized shapes against traditional baseline designs. To this end, we benchmark the best superformula-derived microelements against a square microelement under identical numerical
conditions and target positions. The purpose of this comparison is not to claim
absolute optimality, but rather to assess whether increased geometric complexity
provides a meaningful performance advantage. The same time-reversal phase-only illumination procedure and nanojet metrics are used for all geometries. The square microelement is compared against the top-$10$
candidates from the restricted four-corner superformula sweep, selected by their
averaged composite scores. 

Interestingly, the square microelement achieves a higher composite score,
outperforming the best superformula-derived geometry by approximately $40\%$.
This suggests that simple, highly symmetric shapes can provide robust and
efficient nanojet formation, despite their reduced geometric flexibility. A possible explanation is that the square boundary provides a stable and predictable mapping that is equally efficient over the entire scanning area. In
contrast, more-complex superformula shapes often include additional curvature
and fine-scale features that increase variability. These results indicate that the square microelement offers a
strong and reliable baseline for photonic nanojet steering using time-reversal
phase modulation.

\textbf{Multiple Nanojet Formation.} In the time-reversal framework, a single
entrance-phase profile can be designed to generate multiple 
nanojets simultaneously. Rather than back-propagating from a single desired
focus location, the backward problem can be formulated with several target
positions whose fields are superposed. This produces
one phase-only illumination profile that attempts to reconstruct multiple
localized intensity maxima in the forward propagation.

\begin{figure}[H]
\centering

\begin{subfigure}{0.32\textwidth}
  \includegraphics[
    width=\linewidth,
    trim=1.5cm 0cm 3.0cm 0cm,
    clip
  ]{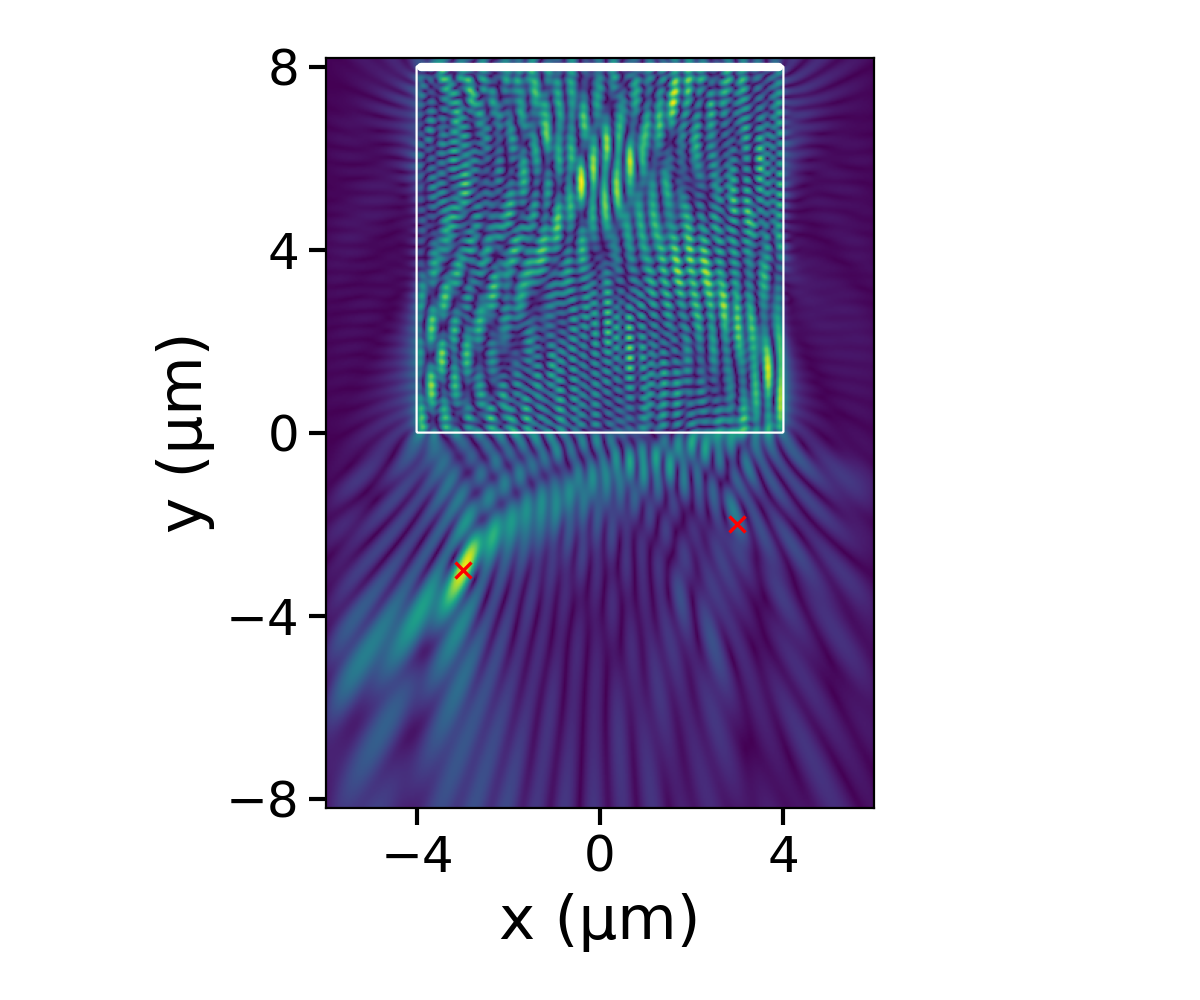}
  \caption{x2}
\end{subfigure}\hfill
\begin{subfigure}{0.32\textwidth}
  \includegraphics[width=\linewidth,
    trim=1.5cm 0cm 3.0cm 0cm,
    clip]{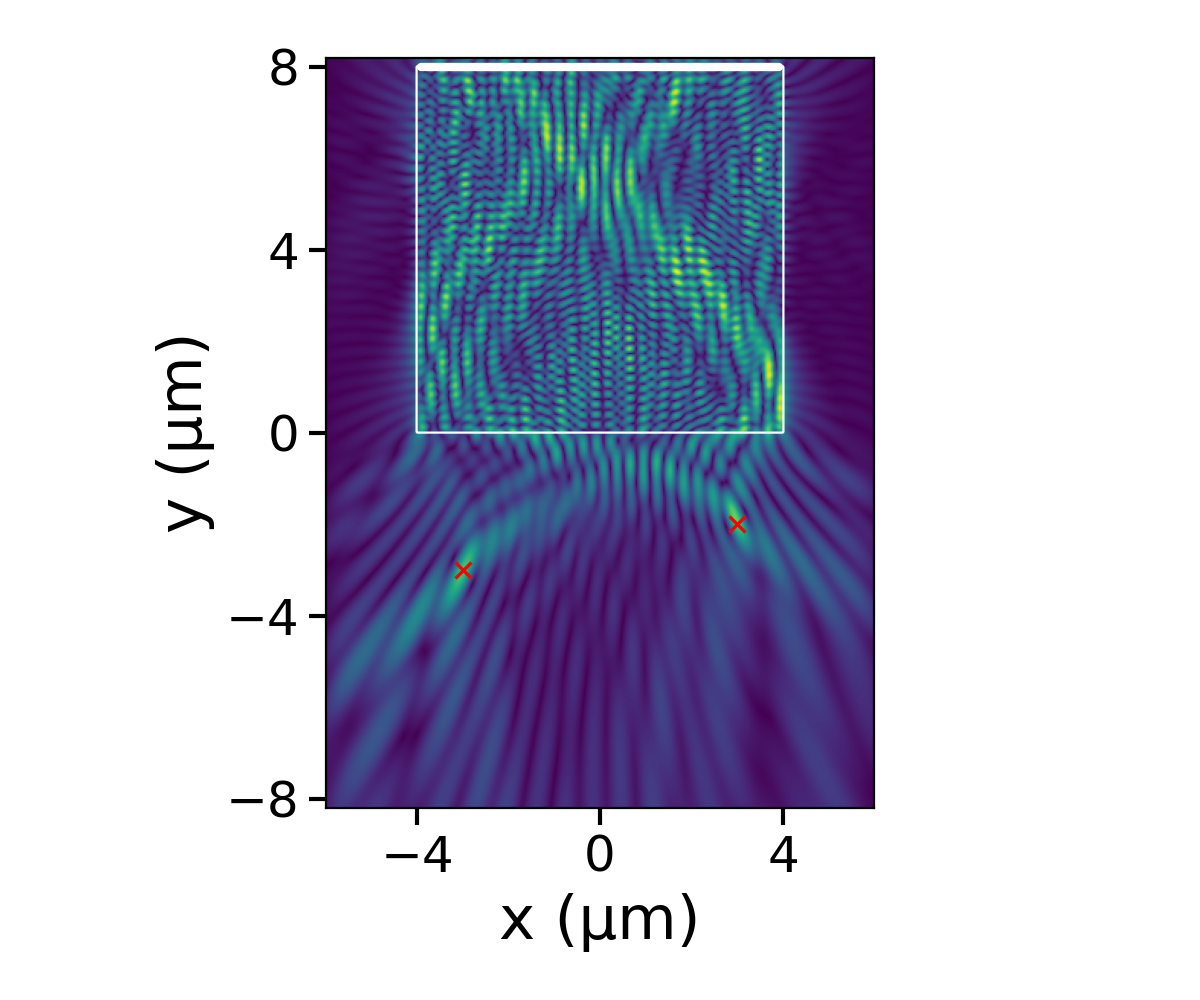}
  \caption{x6}
\end{subfigure}\hfill
\begin{subfigure}{0.32\textwidth}
  \includegraphics[width=\linewidth,
    trim=1.5cm 0cm 3.0cm 0cm,
    clip]{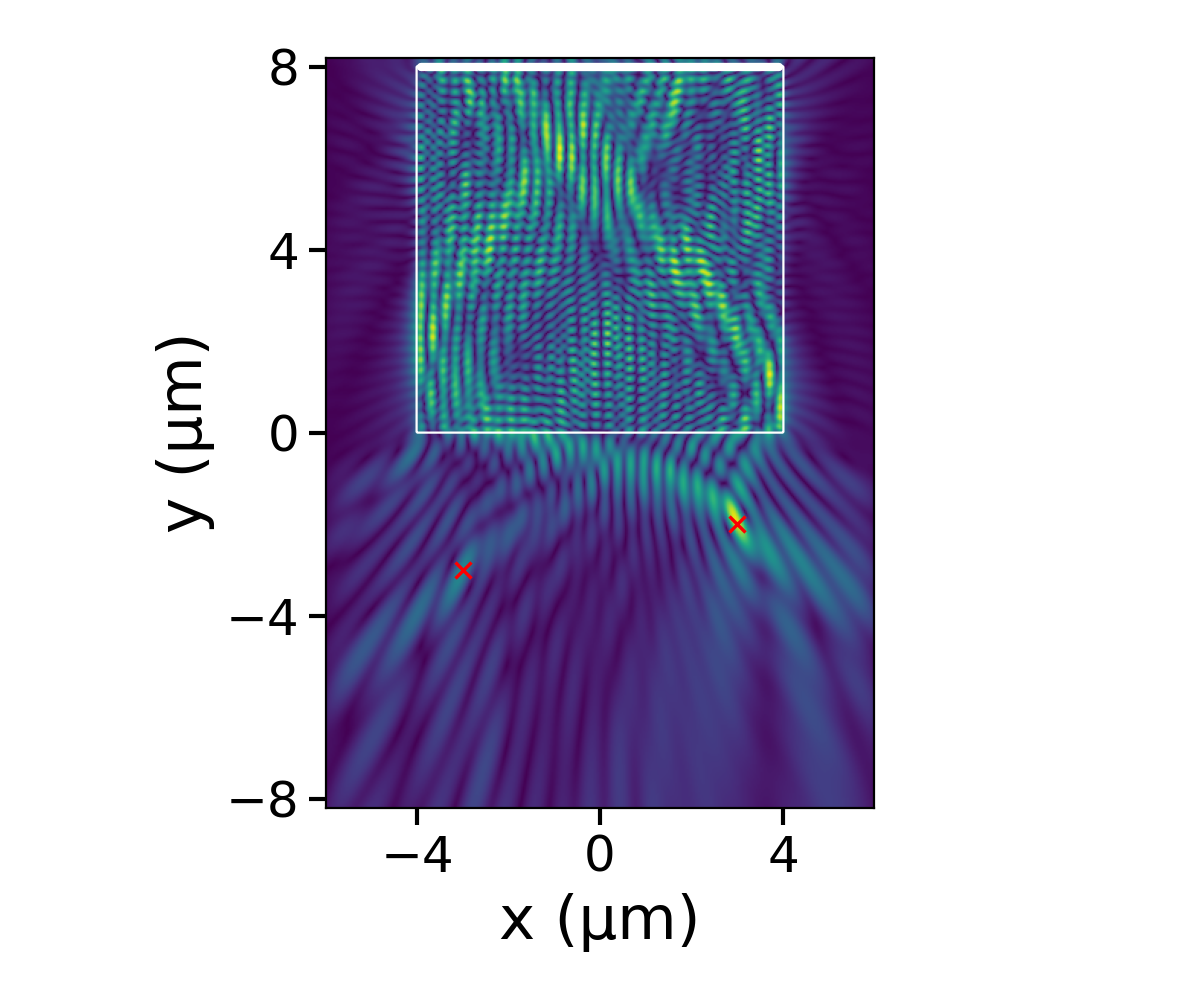}
  \caption{x12}
\end{subfigure}
\caption{Effect of source weighting in the backward solve. The right-hand target source weight is scaled by $\times2$ (a), $\times6$ (b), and $\times12$ (c), showing the transition from left-dominated to balanced to right-dominated reconstruction \cite{Thesis}.}
\label{fig:three_doubles}

\end{figure}

Let the desired nanojet target positions be denoted
$\{\bm{r}_p\}_{p=1}^P$. The backward excitation is constructed as a
superposition of $P$ localized Gaussian sources, each centered at one of the
target locations:

\begin{equation}
S_{\mathrm{back}}(\bm{r})
\;=\;
\sum_{p=1}^{P} w_p
\exp\!\left(
-\frac{\lvert \bm{r}-\bm{r}_p\rvert^2}{2\sigma^2}
\right)
\label{eq:multi_back_source}
\end{equation}

\noindent
where $\sigma$ controls the spatial extent of each source and $w_p$ are optional
weights used to adjust the relative power delivered to each target.  Forward propagation with the calculated phase profile produces an interference pattern that attempts to reconstruct multiple localized maxima near the prescribed targets.
Representative examples of double-nanojet formation are shown in
Fig.~\ref{fig:three_doubles}.

Interestingly, targets located at the same axial depth typically require no
additional weighting to produce nanojets of comparable intensity. However, as
the axial separation increases, progressively larger weights are needed to
compensate for the imbalance. This trend indicates a systematic relationship
between target depth and the power distribution in the backward superposition.
A predictive expression for the optimal weighting is left for future work.

\textbf{Uncertainty quantification.} To assess robustness under realistic fabrication, material, and alignment
errors, we evaluate steering performance with independent perturbations of the
microelement radius, refractive index, and lateral position. We first design a phase profile for the nominal geometry and then reuse it without modification while introducing perturbations on the physical parameters.

We remodel the perturbed variables as
\[
R_{me} = R_0(1+\delta_R), \quad
n_{me} = n_0 + \delta_n, \quad
(x_0,y_0)=(x_0^{(0)},y_0^{(0)})+(\delta_x,\delta_y),
\]
with uniformly sampled tolerances
\[
\delta_n \in [-0.05,0.05], \qquad
\delta_R \in [-0.03,0.03], \qquad
\delta_x,\delta_y \in [-0.10,0.10]~\mu\mathrm{m}.
\]
For each realization, we recompute the forward field using the nominal phase-only illumination and the PNJ metrics are extracted. We quantify the variability by using the mean $\mu$, standard
deviation $\sigma$, and coefficient of variation $CV=\sigma/\mu$ over each set of 15
samples per perturbation class. This is then repeated for each of 6 evenly distributed target locations.

\begin{table}[H]
\centering
\caption{Uncertainty quantification for a square microelement (mean, standard deviation, and coefficient of variation)~\cite{Thesis}.}
\label{tab:uq_cv}
\begin{tabular}{l l c c c}
\toprule
\textbf{Group} & \textbf{Metric} & $\mu$ & $\sigma$ & \textbf{CV} \\
\midrule
$n$
& Waist width $w$ [$\mu$m]        & 0.443 & 0.069 & 0.156 \\
& Axial length $l$ [$\mu$m]        & 2.572 & 0.918 & 0.357 \\
& Peak intensity $I$ & 0.9352 & 0.1120 & 0.120 \\
\midrule
$R$
& Waist width $w$ [$\mu$m]        & 0.443 & 0.075 & 0.169 \\
& Axial length $l$ [$\mu$m]        & 2.550 & 1.006 & 0.395 \\
& Peak intensity $I$ & 0.9443 & 0.0839 & 0.089 \\
\midrule
$xy$
& Waist width $w$ [$\mu$m]        & 0.409 & 0.093 & 0.227 \\
& Axial length $l$ [$\mu$m]        & 2.139 & 1.223 & 0.572 \\
& Peak intensity $I$ & 0.7369 & 0.2373 & 0.322 \\
\bottomrule
\end{tabular}
\end{table}

\noindent
As a consistency check, we use a circular microelement, for which the phase delay
scales as $\phi \sim k_0 nR$. Consequently, perturbations in refractive index
and radius are optically equivalent to first order and yield similar nanojet statistics. The observed agreement in the coefficients of variation verifies the
numerical formulation.

\begin{table}[H]
\centering
\caption{Uncertainty quantification for a circular microelement summary (mean, standard deviation, and coefficient of variation)~\cite{Thesis}.}
\label{tab:uq_cv}
\begin{tabular}{l l c c c}
\toprule
\textbf{Group} & \textbf{Metric} & $\mu$ & $\sigma$ & \textbf{CV} \\
\midrule
$n$
& Waist width $w$ [$\mu$m]        & 0.609 & 0.122 & 0.200 \\
& Axial length $l$ [$\mu$m]        & 4.814 & 1.755 & 0.365 \\
& Peak intensity $I$ & 0.7383 & 0.1559 & 0.211 \\
\midrule
$R$
& Waist width $w$ [$\mu$m]        & 0.604 & 0.118 & 0.195 \\
& Axial length $l$ [$\mu$m]        & 4.793 & 1.626 & 0.339 \\
& Peak intensity $I$ & 0.7298 & 0.1817 & 0.249 \\
\bottomrule
\end{tabular}
\end{table}

\section{Conclusion}\label{sec:conclusion}
We demonstrated numerically a robust, phase-only, time reversal-based steering of photonic nanojets that involves no opto-mechanical intervention and that uses a single fixed high-symmetry microelement. Specifically, we validated wide-range lateral and axial PNJ steering in simulation, as well as the robustness to microelement variability and alignment errors.

\section*{Funding}
This work was funded by the Villum Foundation project no. 58857.

\end{document}